# Collapsing perfect fluid in self-similar five dimensional space-time and cosmic censorship

S. G. Ghosh*
*Department of Mathematics, Science College, Congress Nagar, Nagpur-440 012, India*

S. B. Sarwe[†]
*Department of Mathematics, St. Francis De Sales' College, Seminary Hills, Nagpur-440 006, India*

R. V. Saraykar[‡]
*Department of Mathematics, Nagpur University, Nagpur-440 010, India*

We investigate the occurrence and nature of naked singularities in the gravitational collapse of a self-similar adiabatic perfect fluid in a five dimensional space-time. The naked singularities are found to be gravitationally strong in the sense of Tipler and thus violate the cosmic censorship conjecture.

## I. INTRODUCTION

The early results of Oppenheimer and Snyder [1] on the spherical collapse of a homogeneous dust cloud led to the general concept of trapped surfaces and a black hole. This has motivated the formulations of different cosmic censorship conjectures (CCC) (Penrose [2]). The CCC remains one of the most important unsolved problems in general relativity and gravitation theory today. According to the strong version of the CCC the singularities that appear in gravitational collapse are hidden from all observers. It is weakened to exclude only *globally* naked singularities, i.e., singularities visible to an asymptotic observer as opposed to *locally* naked singularities that are visible to a nearby observer, but may well be hidden behind an event horizon (see [3] for reviews of the CCC). The CCC has as yet no precise mathematical formulation or proof for either version. Consequently, examples that appear to violate the CCC are very important and they are useful tools to study this important issue.

Motivated by the development of superstring and other field theories, there has been a proliferation of articles on higher-dimensional (HD) space-time from the viewpoint of both cosmology [4] and gravitational collapse [5]. In this context, one question that is important and naturally arises is as follows: would the examples of naked singularities in four-dimensional (4D) spherical gravitational collapse go over to higher dimensional (HD) space-time or not? If the answer to this question is yes, then a related question is whether the dimensionality of space-time has any effect on the formation and nature of the singularity. In a recent communication Ghosh and Dadhich [6] employed the Vaidya null radiation collapse scenario to study this effect and showed that qualitatively the situation remains similar to the monotonic shrinkage of a naked singularity window with an increase in dimensions; i.e., an increase in dimensions favors a black hole. A similar situation also occurs in HD dust collapse [7]. The main deficiency of such models is that they neglect pressure, which is likely to diverge at singularities. Also one can argue that the phenomenon of a naked singularity is an artifact of dust or null dust solutions (i.e. one with vanishing pressure). Thus it is obvious that the effect of pressure on the formation of a naked singularity should be taken into account.

In the 4D case, Ori and Piran [8] investigated the collapse of an adiabatic perfect fluid numerically under the assumption of self-similarity and showed that a naked singularity forms, which was followed by analytical discussion [9,10] based on self-similarity. Indeed, the appearance of naked singularities in this type of collapse would appear to be a common feature with soft equation of state [11].

The self-similarity in general relativity is interesting to study for at least three reasons: It generalizes the classical notion of the similarity; it reduces the Einstein field equations to a system of ordinary differential equations; and it provides ''constants of motion'' that may simplify physical interpretation of models. It is therefore interesting to study gravitational collapse of self-similar perfect fluid in 5D space-times from the viewpoint of the CCC.

In this paper we generalize previous studies of 4D spherical self-similar perfect fluid collapse to a five dimensional (5D) space-time and show that gravitational collapse of 5D space-time gives rise to shell focusing naked singularities. This is discussed in Sec. III. The issues of global visibility and strength of the singularities are discussed, respectively, in Sec. IV and Sec. V. In Sec. II we find 5D version of Cahill and Taub [12] spherically symmetric self-similar perfect fluid space-time with adiabatic equation of state. In Sec. VI we show matching of the solution to the outgoing Vaidya solution, which is followed by our discussion and concluding remarks.

We have used units that fix the speed of light and the gravitational constant via $G = c = 1$.

## II. FIVE DIMENSIONAL SELF-SIMILAR SPACE-TIME

A space-time is called self-similar if, by a suitable transformation of coordinates, all metric coefficients and depen-dent variables can be put in the form in which they are functions of a single independent variable, which is dimensionless combination of space and time coordinates. The 5D self-similar spherically symmetric space-time in co-moving coordinates is given by

$$ds^2 = -e^{2\nu}dt^2 + e^{2\psi}dr^2 + r^2 S^2 d\Omega^2 \qquad (1)$$

where $\nu$, $\psi$ and $S$ are functions of the self-similarity variable $X = t/r$, and

$$d\Omega^2 = d\theta^2 + \sin^2\theta d\phi^2 + \sin^2\theta \sin^2\phi d\vartheta^2 \qquad (2)$$

is the metric on an 3-sphere. The stress-energy tensor for a perfect fluid:

$$T^{ab} = (\rho + P)u^a u^b + P g^{ab} \qquad (3)$$

where $u_a = \delta_a^t$ is the 5-dimensional velocity. In the self-similar case, the pressure and energy can be put in the form

$$P = \frac{p(X)}{8\pi r^2}, \quad \rho = \frac{\zeta(X)}{8\pi r^2}. \qquad (4)$$

---

*Author to whom all correspondence should be directed. Electronic address: sgghosh@hotmail.com
[†]Electronic address: sbsarwe@rediffmail.com
[‡]Electronic address: sarayaka@nagpur.dot.net.in

The self-similarity implies the existence of constants of motion along $dX = 0$, which in turn allows the reduction of the Einstein field equations to a set of ordinary differential equations [8]:

$$G_t^t = -\frac{1}{S^2} + \frac{e^{-2\psi}}{S}\left[X^2\ddot{S} - X^2\dot{S}\dot{\psi} + XS\dot{\psi} + \frac{(S-X\dot{S})^2}{S}\right]$$
$$-\frac{e^{-2\nu}}{S}\left(\dot{S}\dot{\psi} + \frac{\dot{S}^2}{S}\right)$$
$$= \frac{-\zeta}{3} \qquad (5)$$

$$G_r^r = -\frac{1}{S^2} + \frac{e^{-2\psi}}{S}\left[-SX\dot{\nu} + X^2\dot{\nu}\dot{S} + \frac{(S-X\dot{S})^2}{S}\right]$$
$$-\frac{e^{-2\nu}}{S}\left(\ddot{S} + \frac{\dot{S}^2}{S} - \dot{\nu}\dot{S}\right)$$
$$= \frac{p}{3} \qquad (6)$$

$$G_\theta^\theta = G_\phi^\phi = G_\vartheta^\vartheta = -\frac{1}{S^2} + \frac{e^{-2\psi}}{S}\left[SX^2\dot{\nu}^2 + 2XS\dot{\psi} - SX^2\dot{\nu}\dot{\psi}\right.$$
$$\left. + 2X^2\dot{\nu}\dot{S} - 2X^2\dot{\psi}\dot{S} + X^2S\ddot{\nu} + 2X^2\ddot{S} + \frac{(S-X\dot{S})^2}{S}\right]$$
$$+ \frac{e^{-2\nu}}{S}\left[S\dot{\nu}\dot{\psi} - S\dot{\psi}^2 + 2\dot{\nu}\dot{S} - 2\dot{\psi}\dot{S} - \frac{\dot{S}^2}{S} - S\ddot{\psi} - 2\ddot{S}\right]$$
$$= p \qquad (7)$$

$$G_r^t = \ddot{S} - \dot{S}\dot{\nu} - \dot{S}\dot{\psi} + \frac{S\dot{\psi}}{X} = 0 \qquad (8)$$

where the overdot denotes the derivative with respect to $X$. We assume that the collapsing fluid obeys an adiabatic equation of state

$$p(X) = \lambda \zeta(X) \qquad (9)$$

where $0 \leq \lambda \leq 1$ is a constant. From the Bianchi identity

$$T^{ab}_{;b} = 0 \qquad (10)$$

it follows that

$$\dot{p} + \frac{2p}{X} = -(\zeta + p)\dot{\nu} \qquad (11)$$

$$\dot{\zeta} = -(\zeta + p)\left[\dot{\psi} + \frac{3\dot{S}}{S}\right]. \qquad (12)$$

Integration of Eqs. (11) and (12), respectively, gives

$$e^{2\nu} = \gamma(\zeta X^2)^{-2\lambda/(1+\lambda)} \qquad (13)$$

$$e^{2\psi} = \alpha(\zeta)^{-2/(1+\lambda)} S^{-6} \qquad (14)$$

where $\alpha$ and $\gamma$ are integration constants. Elimination of $\ddot{S}$ from Eqs. (5) and (6) leads to

$$\left(\frac{\dot{S}}{S}\right)^2 V + \frac{\dot{S}}{S}\left(\frac{\dot{V}}{2} + 3Xe^{2\nu}\right) + e^{2\psi + 2\nu}\left(-\frac{\zeta}{3} - e^{-2\psi} + \frac{1}{S^2}\right) = 0 \qquad (15)$$

and

$$\dot{V} = 2Xe^{2\nu}\left[\frac{1}{3}(\zeta + p)e^{2\psi} - 1\right] = \frac{2}{3}Xe^{2\nu}(H-3) \qquad (16)$$

where the quantities $V$ and $H$ are defined as

$$V(X) = e^{2\psi} - X^2 e^{2\nu}, \quad H = (\zeta + p)e^{2\psi}. \qquad (17)$$

One can also write $H$ as

$$H = 8\pi r^2 e^{2\psi}(T_1^1 - T_0^0). \qquad (18)$$

The matter satisfy the weak energy condition [13] if and only if

$$T_{ab}V^a V^b \geq 0 \qquad (19)$$

for all nonspacelike vector $V^a$. Thus for matter satisfying the weak energy condition, it follows that $H(X) \geq 0$ for all $X$.

## III. THE EXISTENCE OF NAKED SINGULARITIES

In this section, we employ the above solution for investigation of the formation of a black hole or a naked singularity in collapse self-similar adiabatic perfect fluid in a 5D space-time. A self-similar space-time is characterized by the exis-tence of a homothetic Killing vector:

$$\xi^a = r\frac{\partial}{\partial r} + t\frac{\partial}{\partial t} \quad (20)$$

which is given by the Lie derivative

$$\mathcal{L}_\xi g_{ab} = \xi_{a;b} + \xi_{b;a} = 2g_{ab} \quad (21)$$

where $\mathcal{L}$ denotes the Lie derivative. Let $K^a = dx^a/dk$ be the tangent vector to the null geodesics, where $k$ is an affine parameter. Then

$$g_{ab}K^aK^b = 0. \quad (22)$$

It follows that along null geodesics, we have

$$\xi^a K_a = C \quad (23)$$

where $C$ is a constant. From the above algebraic equation and the null condition (22), we get

$$re^{2\psi}K^r - te^{2\nu}K^t = C \quad (24)$$

$$-e^{2\nu}(K^t)^2 + e^{2\psi}(K^r)^2 = 0. \quad (25)$$

Solving the above equations yields the following exact expressions for $K^t$ and $K^r$:

$$K^t = \frac{C[X \pm e^{2\psi}Q]}{r[e^{2\psi} - e^{2\nu}X^2]} \quad (26)$$

$$K^r = \frac{C[1 \pm Xe^{2\nu}Q]}{r[e^{2\psi} - e^{2\nu}X^2]} \quad (27)$$

where $Q = \sqrt{e^{-2\psi-2\nu}} > 0$. To study the nature of the singularity, we employ the technique developed by Dwivedi and Joshi [9] by making necessary changes for the 5D case. Radial ($\theta$, $\phi$ and $\vartheta$ = const) null geodesics, by virtue of Eqs. (26) and (27), satisfy

$$\frac{dt}{dr} = \frac{X \pm e^{2\psi}Q}{1 \pm Xe^{2\nu}Q}. \quad (28)$$

At this point, we note that a curvature singularity forms at the origin ($t=0, r=0$), where the physical quantities like density diverge. This divergence of the density in this singularity results also in a divergence of curvature scalars there. The nature of the singularity (a naked singularity or a black hole) can be characterized by the existence of radial null geodesics emerging from the singularity. The singularity is at least locally naked if there exist such geodesics, and if no such geodesics exist it is a black hole. If the singularity is naked, then there exists a real and positive value of $X_0$ as a solution to the algebraic equation [3]

$$X_0 = \lim_{t \to 0 \, r \to 0} X = \lim_{t \to 0 \, r \to 0} \frac{t}{r} = \lim_{t \to 0 \, r \to 0} \frac{dt}{dr}. \quad (29)$$

Using Eq. (28) and L'Hôpital's rule we can derive the following equation:

$$V(X_0)Q(X_0) = 0. \quad (30)$$

Since $Q > 0$, this implies that

$$V(X_0) = 0. \quad (31)$$

This algebraic equation governs the behavior of the tangent near the singular points. The central shell focusing is at least locally naked (for brevity we address it as naked throughout this paper) if Eq. (31) admits one or more positive roots. The values of the roots give the tangents of the escaping geodesics near the singularity. The smallest value of $X_0$, say $X_0^s$, corresponds to the earliest ray escaping from the singularity, which is called the Cauchy horizon of the space-time and there is no solution in the region $X < X_0^s$. Thus in the absence of a positive root to Eq. (31), the central singularity is not naked because there is no outgoing future directed null geodesics emanating from the singularity.

Next, we ask the question of when this will be realized in terms of the parameter in self-similar field equations. This can be achieved by further analysis of the self-similar field equations. To this end, we define two new functions $y = X^\beta$ and $U^2 = e^{2\psi - 2\nu}/X^2 = y^{-3}\zeta^{-3\beta}S^{-6}$ where $0 \leq \lambda \leq 1$, $\delta = 1 + \lambda$ and $\beta = 2(1-\lambda)/3(1+\lambda)$. Inserting these transformations into Eqs. (15) and (16) yields

$$\beta y \frac{\zeta'}{\zeta} = \frac{1}{U^2 - \lambda}\left(2\lambda - 3\delta\beta y U^2 \frac{S'}{S} - \frac{1}{3}\delta^2 y^3 \zeta^{3\beta/2} U^2\right) \quad (32)$$

and

$$\left(\frac{S'}{S}\right)^2 \beta^2 y^2 (U^2 - 1) + \left(\frac{S'}{S}\right)\beta y \left[2 + \frac{1}{3}\delta y^3 \zeta^{3\beta/2} U^2\right]$$
$$- \left[1 + \left(1 - \frac{3}{\zeta S^2}\right)\frac{1}{3}y^3 \zeta^{3\beta/2} U^2\right] = 0 \quad (33)$$

where the dash denotes differentiation with respect to $y$. The scale invariant quantity U represents velocity of the fluid relative to the hypersurface $X$ = const. The case, where $U = 1$, is of special interest to us as this corresponds to a situation that the space-time allows a naked singularity at the center, i.e., nonspacelike geodesic terminate at the singularity in the past [9]. Thus we are interested in the values of different parameters for the solution, which take into account the case $U = 1$ for some $X > 0$. We analyze the above differential equation near the point $y = y_0 = y(X_0)$ with the condition that $U(y_0) = (\zeta_0 X_0)^{-3\beta}S_0^{-6} = 1$.

Let us write

$$\zeta(X) = \zeta_0 + \zeta_0 \sum_{k=1}^{\infty} \zeta_k (y - y_0)^k \quad (34)$$

and

$$S(X) = S_0 + S_0 \sum_{k=1}^{\infty} S_k (y - y_0)^k. \tag{35}$$

On using Eqs. (34) and (35), Eqs. (32) and (33) take, respectively, the forms

$$\zeta_1 = \frac{1}{\beta y_0 (1-\lambda)} \left( 2\lambda - 3\beta \delta y_0 S_1 - \frac{1}{3} \delta^2 y_0^3 \zeta_0^{3\beta/2} \right) \tag{36}$$

and

$$\beta y_0 S_1 = \frac{3}{6 + \delta y_0^3 \zeta_0^{3\beta/2}} \left[ 1 + \frac{1}{3} \left( 1 - \frac{3}{\zeta_0 S_0^2} \right) y_0^3 \zeta_0^{3\beta/2} \right]. \tag{37}$$

Here $\zeta'(y_0) = \zeta_0 \zeta_1$ and $S'(y_0) = S_0 S_1$. Eliminating $S_1$ and $S_0$ from the above equations we obtain

$$Y^6 + (mn - 9n)Y^4 + (6\delta - 6\lambda + 9)Y^3 + 6\delta mn Y + 27\delta^2$$
$$- 36\lambda \delta = 0 \tag{38}$$

where

$$Y = \delta^{2/3} \zeta_0^{\beta/2} y_0, \quad m = \frac{\beta(1-\lambda)\zeta_1}{\zeta_0^{\beta-1}}, \quad n = \frac{3\zeta_0^{(\beta-2)/2}}{\delta^{2/3}}.$$

This algebraic equation ultimately decides the final fate of the collapse. In general the existence of real positive roots of the above sixth degree algebraic equation will put a limitation on the physical parameters $\zeta_0$ and $\zeta_1$. Thus the existence of real positive roots of $V(X) = 0$ (and, hence existence of naked singularity) is characterized by the values of physical parameters $\zeta_0$ and $\zeta_1$. It is easy to check that the above equation can admit at the most four real positive roots. In the similar situation of 4D, one gets a quartic equation.

## IV. GLOBAL VISIBILITY

A naked singularity can be considered as a physically significant singularity, if it can escape from singularity to far away observers for a finite period of time. The analysis to show this is similar to that of the 4D case as discussed in [9]. Details of the mathematical description can be found in [9].

The singularities are visible for a finite period of time, if infinity of integral curves escapes from singularity. To see this, we write the equation of geodesics in the form $r = r(X)$. Using Eqs. (26) and (27) we obtain

$$\frac{dX}{dr} = \frac{V(X)Q(X)}{r[1 + Xe^{2\nu}Q]}. \tag{39}$$

Integration of the above yields the equation of geodesics curve in $(X, r)$ plane, which can be written as

$$r = D \exp \left[ \int \frac{[1 + Xe^{2\nu}Q]}{V(X)Q(X)} dX \right] \tag{40}$$

where $D$ is an integration constant that labels different geodesics. We have demonstrated that for a singularity to be naked, $V(X) = 0$ must have at least one real positive root. Let $X = X_0$ be a simple root of $V(X) = 0$. We could then decompose V(X) as $V(X) = \frac{2}{3}(X - X_0)X_0 e^{2\nu(X_0)}[H_0 - 3]$. Inserting this expression of $V(X)$ in Eq. (40) yields

$$r = D(X - X_0)^{3/[H_0 - 3]} \tag{41}$$

where $H_0 = H(X_0)$. Thus, when $H_0 > 3$, an infinity of integral curves with tangent $X = X_0$ would escape from singularity and hence singularity would be visible for a finite period of time. In the corresponding 4D case integral curves meet the singularity in the past if $H_0 > 2$ [9]. Further, as in the 4D case, it can be shown that an infinity of integral curves will escape from singularity if $0 < H_0 < \infty$. It is seen that $H_0 > 0$ holds only if the weak energy condition is true. Thus we have shown that infinity of integral curves would escape the singularity provided the weak energy condition is satisfied.

The naked singularities discussed in the previous section are locally naked, i.e., invisible to an asymptotic observer. Next, we briefly describe when geodesics can reach a far away observer. This follows from the analysis of Eq. (40), which can be rearranged as

$$r = D \exp \left[ \int \frac{[1 + Xe^{2\nu}Q]}{(X - a_1)g(X)} dX - \int \frac{[1 + Xe^{2\nu}Q]}{(X - a_2)g(X)} dX \right] \tag{42}$$

where

$$g(X) = \frac{(a_1 - a_2)V(X)Q(X)}{(X - a_1)(X - a_2)}$$

where $a_1$ and $a_2$ (with $a_1 > a_2$) are two consecutive roots of the equation $V(X) = 0$. Thus all integral curves can meet singularity in past with tangent $X = a_1$ and $r = \infty$ can be realized in the future along the same integral curves at $X = a_2$. Thus the singularities will be globally naked and an infinity of curves would emanate from singularity to reach a distant observer.

## V. STRENGTH OF NAKED SINGULARITY

Finally, we need to determine the curvature strength of the naked singularity, which is an important aspect of a singularity [14]. A singularity is gravitationally strong or simply strong if volume elements get crushed to zero dimensions at the singularity, and weak otherwise. It is widely believed that a space-time does not admit an extension through a singularity if it is a strong curvature singularity in the sense of Tipler [15]. A necessary and sufficient condition criterion for a singularity to be strong has been given by Clarke and Królak [16] that for at least one nonspacelike geodesic with affine parameter $k$, in the limiting approach to the singularity, we must have

$$\lim_{k \to 0} k^2 \psi = \lim_{k \to 0} k^2 R_{ab} K^a K^b > 0 \tag{43}$$

where $R_{ab}$ is the Ricci tensor. Our purpose here is to investigate the above condition along future directed radial null geodesics that emanate from the naked singularity. Equation (43) can be expressed as

$$\lim_{k \to 0} k^2 \psi = \lim_{k \to 0} k^2 \frac{(\zeta + p)C^2 e^{2\nu}[X + e^{2\psi}Q]^2}{r^4[e^{2\psi} - X^2 e^{2\nu}]^2}. \quad (44)$$

Using Eqs. (26),(27), and L'Hospitals rule, Eq. (44) turns out to be

$$\lim_{k \to 0} k^2 \psi = \frac{9H_0}{(H_0 + 3)^2} > 0. \quad (45)$$

Thus along radial null geodesics the strong curvature condition is satisfied if $H_0 > 0$, which is also a necessary condition for the energy condition. Thus it follows that singularities are gravitationally strong if the weak energy condition is satisfied.

## VI. MATCHING WITH OUTGOING FIVE DIMENSIONAL VAIDYA SOLUTION

In order to study the collapse of a finite spherical body, we have to match the solution along the timelike surface to a 5D static space-time. We consider a spherical surface with its motion described by a timelike four space $\Sigma$, which divides space-times into interior and exterior manifolds. We shall first cut the space-time along the timelike hypersurface, and then join the internal part with the outgoing 5D Vaidya solution. The metric on the whole space-time can be written in the form

$$ds^2 = \begin{cases} -e^{2\nu}dt^2 + e^{2\psi}dr^2 + r^2 S^2 d\Omega^2, & r \leq r_\Sigma, \\ -\left(1 - \frac{m(v)}{\mathbf{r}^2}\right)dv^2 - 2dv d\mathbf{r} + \mathbf{r}^2 d\Omega^2, & r \geq r_\Sigma. \end{cases} \quad (46)$$

The metric on the hypersurface $r = r_\Sigma$ is given by

$$ds^2 = -d\tau^2 + \mathcal{R}^2(\tau)d\Omega^2. \quad (47)$$

For the junction conditions we suitably modify the approach given in [17,18] for our 5D case. Hence we have to demand

$$(ds_-^2)_\Sigma = (ds_+^2)_\Sigma = (ds^2)_\Sigma. \quad (48)$$

The second junction condition is obtained by requiring the continuity of the extrinsic curvature of $\Sigma$ across the boundary. This yields

$$K_{ij}^- = K_{ij}^+ \quad (49)$$

where $K_{ij}^\pm$ is extrinsic curvature to $\Sigma$, given by

$$K_{ij}^\pm = -n_{\alpha\Delta}^\pm \frac{\partial^2 x_\pm^{\alpha\Delta}}{\partial \xi^i \partial \xi^j} - n_{\alpha\Delta}^\pm \Gamma_{\beta\Delta\gamma\Delta}^{\alpha\Delta} \frac{\partial x_\pm^{\beta\Delta}}{\partial \xi^i} \frac{\partial x_\pm^{\gamma\Delta}}{\partial \xi^j} \quad (50)$$

and where $\Gamma_{\beta\Delta\gamma\Delta}^{\alpha\Delta}$ are Christoffel symbols, $n_{\alpha\Delta}^\pm$ the unit normal vectors to $\Sigma$, $x^{\alpha\Delta}$ are the coordinates of the interior and exterior space-time and $\xi^i$ are the coordinates that define $\Sigma$. From the junction condition (48) we obtain

$$\frac{dt}{d\tau} = \frac{1}{e^{\nu(r_\Sigma, t)}} \quad (51)$$

$$r_\Sigma S(r_\Sigma, t) = \mathbf{r}(\tau) \quad (52)$$

$$\left(\frac{dv}{d\tau}\right)_\Sigma^{-2} = \left(1 - \frac{m(v)}{\mathbf{r}^2} + 2\frac{d\mathbf{r}}{dv}\right)_\Sigma. \quad (53)$$

The nonvanishing components of the intrinsic curvature $K_{ij}$ of $\Sigma$ can be calculated and the result is

$$K_{\tau\tau}^- = (-e^{-\psi}\nu_r)_\Sigma \quad (54a)$$

$$K_{\theta\theta}^- = [e^{-\psi}rS(S + rS_r)]_\Sigma \quad (54b)$$

$$K_{\tau\tau}^+ = \left[\frac{d^2v}{d\tau^2}\left(\frac{dv}{d\tau}\right)^{-1} - \left(\frac{dv}{d\tau}\right)\frac{m(v)}{\mathbf{r}^3}\right]_\Sigma \quad (54c)$$

$$K_{\theta\theta}^+ = \left[\mathbf{r}\frac{d\mathbf{r}}{d\tau} + \left(\frac{dv}{d\tau}\right)\left(1 - \frac{m(v)}{\mathbf{r}^2}\right)\mathbf{r}\right]_\Sigma \quad (54d)$$

$$K_{\phi\phi}^\pm = \sin^2\theta K_{\theta\theta}^\pm \quad (54e)$$

$$K_{\vartheta\vartheta}^\pm = \sin^2\phi K_{\phi\phi}^\pm \quad (54f)$$

where subscripts $r$ and $t$ denote partial derivative with respect to $r$ and $t$ respectively. The unit normal to the $\Sigma$ are given by

$$n_{\alpha\Delta}^- = (0, e^{\psi(r_\Sigma, t)}, 0, 0, 0) \quad (55)$$

$$n_{\alpha\Delta}^+ = \left(1 - \frac{m(v)}{\mathbf{r}^2} + 2\frac{d\mathbf{r}}{dv}\right)^{-1/2}\left(-\frac{d\mathbf{r}}{dv}, 1, 0, 0, 0\right). \quad (56)$$

From Eqs. (49), (54b) and (54d) we have

$$\left[\left(\frac{dv}{d\tau}\right)\left(1 - \frac{m(v)}{\mathbf{r}^2}\right)\mathbf{r} + \mathbf{r}\frac{d\mathbf{r}}{d\tau}\right]_\Sigma = [e^{-\psi}rS(S + rS_r)]_\Sigma. \quad (57)$$

With the help of Eqs. (51), (52) and (53), we can write Eq. (57) as

$$m = r^2 S^2\left[1 + \frac{rS_t^2}{e^{2\nu}} - \frac{(S + rS_r)^2}{e^{2\psi}}\right]_\Sigma \quad (58)$$

which is the total energy entrapped inside the surface $\Sigma$. From Eqs. (54a) and (54c), using Eq. (51), we have

$$\left[\frac{d^2v}{d\tau^2}\left(\frac{dv}{d\tau}\right)^{-1} - \left(\frac{dv}{d\tau}\right)\frac{m(v)}{\mathbf{r}^3}\right]_\Sigma = -(e^{-\psi}\nu_r)_\Sigma. \quad (59)$$

Substituting Eqs. (51), (52) and (58) into Eq. (57) we can write

$$\left(\frac{dv}{d\tau}\right)_\Sigma = \left[\frac{S+rS_r}{e^\psi} + \frac{rS_t}{e^\nu}\right]_\Sigma^{-1}. \quad (60)$$

Differentiating Eq. (60) with respect to $\tau$ and using Eq. (58), we can rewrite Eq. (59) as

$$-\left(\frac{\nu_r}{e^\psi}\right)_\Sigma = \left[\left[-\frac{r}{e^\psi}S_{tr} + \frac{\psi_t(S+rS_r)}{e^\psi} + \frac{r\nu_t S_t}{e^\nu} - \frac{rS_{tt}}{e^\nu} - \frac{rS_t^2}{e^\nu S}\right.\right.$$
$$\left.\left. - \frac{S_t}{e^\psi} + \frac{e^\nu}{rS}\left(\frac{(S+rS_r)^2}{e^{2\psi}} - 1\right)\right]\right.$$
$$\left.\times\left[\frac{S+rS_r}{e^\psi} + \frac{rS_t}{e^\nu}\right]^{-1}\frac{1}{e^\nu}\right]_\Sigma. \quad (61)$$

Next we translate the above equation in terms of $X = t/r$, which on the surface $\Sigma$ is given by

$$-\frac{1}{S^2} + \frac{e^{-2\psi}}{S}\left[-SX\dot\nu + X^2\dot\nu\dot S + \frac{(S-X\dot S)^2}{S}\right] - \frac{e^{-2\nu}}{S}$$
$$\times\left(\ddot S + \frac{\dot S^2}{S} - \dot\nu\dot S\right)$$
$$= \frac{e^{-\psi-\nu}(-X)}{S}\left[\ddot S - \dot S\dot\nu - \dot S\dot\psi + \frac{S\dot\psi}{X}\right]. \quad (62)$$

Comparing Eq. (62) with Eqs. (6) and (8), we can finally write

$$(P)_\Sigma = 0. \quad (63)$$

Equation (63) shows that the pressure will vanish at the boundary which implies radiation cannot exist and exterior space-time is the 5D Schwarzschild space-time.

## VII. DISCUSSION AND CONCLUDING REMARKS

The gravitational collapse of a spherically symmetric self-similar adiabatic perfect fluid in a 5D space-time is studied. The idea for space-time to extend from four to five dimensions was introduced by Kaluza and Klein [19] to unify gravity and electromagnetism. Five dimensional (5D) space-time is particularly more relevant because both 10D and 11D supergravity theories yield solutions where a 5D space-time results after dimensional reduction [20]. Hence, we have confined ourselves to the 5D case.

In the analogous 4D case, both analytical [9] and numerical results [11] indicate the critical behavior governing the formation of black holes or naked singularities. For further progress toward an understanding of spherical collapse, from the viewpoint of the CCC, one would like to know the effect of extra dimensions on the existence of a naked singularity. The relevant questions would be, for instance, whether such solutions remain naked with the introduction of extra dimensions. Do they always become covered? Does the nature of the singularity change? Our analysis shows that none of the above hold. Indeed, the gravitational collapse of a perfect fluid in a 5D self-similar spherically symmetric space-time leads to a strong curvature naked singularity.

We found that 5D spherically symmetric self-similar space-times admits globally strong curvature naked singularities provided the algebraic equation $V(X)=0$ has real positive roots. We have shown that it is the weak energy condition that is necessary for visibility of singularities for a finite period of time and for singularities to be gravitationally strong.

For completeness, we have matched the solution to 5D Vaidya pure radiation solutions so that the resulting solution will represent the collapse of a star.

This generalizes the previous studies of spherical gravitational collapse in 4D to 5D space-times. The formation of these naked singularities violates the CCC. We think that the results obtained here have some interest in the sense that they do offer the opportunity to explore properties associated with naked singularities, which may be crucial in our understanding of this important problem. We restricted our analysis to radial null geodesics; there is no reason to believe that it cannot be extended to nonradial geodesics and other causal curves.

Finally, the result obtained would also be relevant in the context of superstring theory which is often said to be the next "theory of everything," and for an interpretation of how critical behavior depends on the dimensionality of the space-time.

### ACKNOWLEDGMENTS

One of the authors (S.G.G.) would like to thank IUCAA, Pune for hospitality while this work was done.